\documentclass[11pt]{article}
\usepackage[dvips]{graphicx}
\usepackage{epsfig}
\usepackage{amssymb}
\usepackage{color}
\usepackage[left]{lineno}

\begin{document}
\centerline{\LARGE EUROPEAN ORGANIZATION FOR NUCLEAR RESEARCH}
%
%
\vspace{10mm} {\flushright{
CERN-PH-EP-2013-197 \\
17 October 2013\\
Revised version:\\14 January 2014\\
}}
%
%
\vspace{10mm}

\begin{center}
\boldmath
{\bf {\Large \boldmath{A new measurement of the $K^\pm\to\pi^\pm\gamma\gamma$ decay\\ at the NA48/2 experiment}}}
\unboldmath
\end{center}
\begin{center}
{\Large The NA48/2 collaboration}\\
\end{center}

\begin{abstract}
The NA48/2 experiment at CERN collected two data samples with minimum bias trigger conditions in 2003 and 2004. A measurement of the rate and dynamic properties of the rare decay $K^\pm\to\pi^\pm\gamma\gamma$ from these data sets based on 149 decay candidates with an estimated background of $15.5\pm0.7$ events is reported. The model-independent branching ratio in the kinematic range $z=(m_{\gamma\gamma}/m_K)^2>0.2$ is measured to be ${\cal B}_{\rm MI}(z>0.2) = (0.877 \pm 0.089) \times 10^{-6}$, and the branching ratio in the full kinematic range assuming a particular Chiral Perturbation Theory description to be ${\cal B}(K_{\pi\gamma\gamma}) = (0.910 \pm 0.075) \times 10^{-6}$.
\end{abstract}

\begin{center}
{\it Accepted for publication by Physics Letters B}
\end{center}

\newpage
\begin{center}
{\Large The NA48/2 Collaboration}\\
\vspace{2mm}
 J.R.~Batley,
 G.~Kalmus,
 C.~Lazzeroni$\,$\footnotemark[1]$^,$\footnotemark[2],
 D.J.~Munday,
 M.W.~Slater$\,$\footnotemark[1],
 S.A.~Wotton \\
{\em \small Cavendish Laboratory, University of Cambridge,
Cambridge, CB3 0HE, UK$\,$\footnotemark[3]} \\[0.2cm]
 R.~Arcidiacono$\,$\footnotemark[4],
 G.~Bocquet,
 N.~Cabibbo$\,$\footnotemark[5],
 A.~Ceccucci,
 D.~Cundy$\,$\footnotemark[6],
 V.~Falaleev,
 M.~Fidecaro,
 L.~Gatignon,
 A.~Gonidec,
 W.~Kubischta,
 A.~Norton$\,$\footnotemark[7],
 A.~Maier,\\
 M.~Patel$\,$\footnotemark[8],
 A.~Peters\\
{\em \small CERN, CH-1211 Gen\`eve 23, Switzerland} \\[0.2cm]
 S.~Balev$\,$\footnotemark[5],
 P.L.~Frabetti,
 E.~Gersabeck$\,$\footnotemark[10],
 E.~Goudzovski$\,$\renewcommand{\thefootnote}{\fnsymbol{footnote}}%
\footnotemark[1]\renewcommand{\thefootnote}{\arabic{footnote}}$^,$\footnotemark[1]$^,$\footnotemark[2],
 P.~Hristov$\,$\footnotemark[9],
 V.~Kekelidze,
 V.~Kozhuharov$\,$\footnotemark[11],
 L.~Litov$\,$\footnotemark[12],
 D.~Madigozhin,
 N.~Molokanova,
 I.~Polenkevich,
 Yu.~Potrebenikov,
 S.~Stoynev$\,$\footnotemark[13],
 A.~Zinchenko \\
{\em \small Joint Institute for Nuclear Research, 141980 Dubna (MO), Russia} \\[0.2cm]
 E.~Monnier$\,$\footnotemark[14],
 E.~Swallow,
 R.~Winston$\,$\footnotemark[15]\\
{\em \small The Enrico Fermi Institute, The University of Chicago,
Chicago, IL 60126, USA}\\[0.2cm]
 P.~Rubin$\,$\footnotemark[16],
 A.~Walker \\
{\em \small Department of Physics and Astronomy, University of
Edinburgh, Edinburgh, EH9 3JZ, UK} \\[0.2cm]
 W.~Baldini,
 A.~Cotta Ramusino,
 P.~Dalpiaz,
 C.~Damiani,
 M.~Fiorini,
 A.~Gianoli, \\
 M.~Martini,
 F.~Petrucci,
 M.~Savri\'e,
 M.~Scarpa,
 H.~Wahl \\
 {\em \small Dipartimento di Fisica e Scienze della Terra dell'Universit\`a e Sezione
dell'INFN di Ferrara, \\ I-44122 Ferrara, Italy} \\[0.2cm]
 A.~Bizzeti$\,$\footnotemark[17],
 M.~Lenti,
 M.~Veltri$\,$\footnotemark[18] \\
{\em \small Sezione dell'INFN di Firenze, I-50019 Sesto Fiorentino, Italy} \\[0.2cm]
 M.~Calvetti,
 E.~Celeghini,
 E.~Iacopini,
 G.~Ruggiero$\,$\footnotemark[9] \\
{\em \small Dipartimento di Fisica dell'Universit\`a e Sezione
dell'INFN di Firenze, I-50019 Sesto Fiorentino, Italy} \\[0.2cm]
 M.~Behler,
 K.~Eppard,
 K.~Kleinknecht,
 P.~Marouelli,
 L.~Masetti,
 U.~Moosbrugger,\\
 C.~Morales Morales$\,$\footnotemark[19],
 B.~Renk,
 M.~Wache,
 R.~Wanke,
 A.~Winhart$\,$\footnotemark[1]\\
{\em \small Institut f\"ur Physik, Universit\"at Mainz, D-55099 Mainz, Germany$\,$\footnotemark[20]} \\[0.2cm]
 D.~Coward$\,$\footnotemark[21],
 A.~Dabrowski$\,$\footnotemark[9],
 T.~Fonseca Martin,
 M.~Shieh,
 M.~Szleper,\\
 M.~Velasco,
 M.D.~Wood$\,$\footnotemark[21] \\
{\em \small Department of Physics and Astronomy, Northwestern
University, Evanston, IL 60208, USA}\\[0.2cm]
 P.~Cenci,
 M.~Pepe,
 M.C.~Petrucci \\
{\em \small Sezione dell'INFN di Perugia, I-06100 Perugia, Italy} \\[0.2cm]
 G.~Anzivino,
 E.~Imbergamo,
 A.~Nappi$\,$\footnotemark[5],
 M.~Piccini,
 M.~Raggi$\,$\footnotemark[11],
 M.~Valdata-Nappi \\
{\em \small Dipartimento di Fisica dell'Universit\`a e
Sezione dell'INFN di Perugia, I-06100 Perugia, Italy} \\[0.2cm]
 C.~Cerri,
 R.~Fantechi \\
{\em Sezione dell'INFN di Pisa, I-56100 Pisa, Italy} \\[0.2cm]
 G.~Collazuol$\,$\footnotemark[22],
 L.~DiLella,
 G.~Lamanna,
 I.~Mannelli,
 A.~Michetti \\
{\em Scuola Normale Superiore e Sezione dell'INFN di Pisa, I-56100
Pisa, Italy} \\[0.2cm]
 F.~Costantini,
 N.~Doble,
 L.~Fiorini$\,$\footnotemark[23],
 S.~Giudici,
 G.~Pierazzini$\,$\footnotemark[5],
 M.~Sozzi,
 S.~Venditti$\,$\footnotemark[9]\\
{\em Dipartimento di Fisica dell'Universit\`a e Sezione dell'INFN di
Pisa, I-56100 Pisa, Italy} \\[0.2cm]
\newpage
 B.~Bloch-Devaux$\,$\footnotemark[24],
 C.~Cheshkov$\,$\footnotemark[25],
 J.B.~Ch\`eze,
 M.~De Beer,
 J.~Derr\'e,
 G.~Marel,
 E.~Mazzucato,
 B.~Peyaud,
 B.~Vallage \\
{\em \small DSM/IRFU -- CEA Saclay, F-91191 Gif-sur-Yvette, France} \\[0.2cm]
 M.~Holder,
 M.~Ziolkowski \\
{\em \small Fachbereich Physik, Universit\"at Siegen, D-57068 Siegen, Germany$\,$\footnotemark[26]} \\[0.2cm]
 C.~Biino,
 N.~Cartiglia,
 F.~Marchetto \\
{\em \small Sezione dell'INFN di Torino, I-10125 Torino, Italy} \\[0.2cm]
 S.~Bifani$\,$\footnotemark[1],
 M.~Clemencic$\,$\footnotemark[9],
 S.~Goy Lopez$\,$\footnotemark[27]\\
{\em \small Dipartimento di Fisica dell'Universit\`a e
Sezione dell'INFN di Torino, I-10125 Torino, Italy} \\[0.2cm]
 H.~Dibon,
 M.~Jeitler,
 M.~Markytan,
 I.~Mikulec,
 G.~Neuhofer,
 L.~Widhalm$\,$\footnotemark[5] \\
{\em \small \"Osterreichische Akademie der Wissenschaften, Institut
f\"ur Hochenergiephysik,\\ A-10560 Wien, Austria$\,$\footnotemark[28]} \\[0.5cm]
\end{center}

%
\renewcommand{\thefootnote}{\fnsymbol{footnote}}
\footnotetext[1]{Corresponding author, email: eg@hep.ph.bham.ac.uk}
\renewcommand{\thefootnote}{\arabic{footnote}}
\footnotetext[1]{Present address: School of Physics and Astronomy, University of Birmingham, Birmingham, B15 2TT, UK}
\footnotetext[2]{Supported by a Royal Society University Research Fellowship}
\footnotetext[3]{Funded by the UK Particle Physics and Astronomy Research Council}
\footnotetext[4]{Present address: Universit\`a degli Studi del Piemonte Orientale, I-13100 Vercelli, Italy}
\footnotetext[5]{Deceased}
\footnotetext[6]{Present address: Istituto di Cosmogeofisica del CNR di Torino, I-10133 Torino, Italy}
\footnotetext[7]{Present address: Dipartimento di Fisica e Scienze della Terra dell'Universit\`a e Sezione dell'INFN di Ferrara, I-44122 Ferrara, Italy}
\footnotetext[8]{Present address: Department of Physics, Imperial College, London, SW7 2BW, UK}
\footnotetext[9]{Present address: CERN, CH-1211 Gen\`eve 23, Switzerland}
\footnotetext[10]{Present address: Ruprecht-Karls-Universit\"{a}t Heidelberg, D-69120 Heidelberg, Germany}
\footnotetext[11]{Present address: Laboratori Nazionali di Frascati, I-00044 Frascati, Italy}
\footnotetext[12]{Present address: Faculty of Physics, University of Sofia ``St. Kl. Ohridski'', 1164 Sofia, Bulgaria, funded by the Bulgarian National Science Fund under contract DID02-22}
\footnotetext[13]{Present address: Northwestern University, Evanston, IL 60208, USA}
\footnotetext[14]{Present address: Universit\'e de la M\'editerran\'ee, F-13013  Marseille, France}
\footnotetext[15]{Present address: University of California, Merced, CA 95344, USA}
\footnotetext[16]{Present address: Department of Physics and Astronomy, George Mason University, Fairfax, VA 22030, USA}
\footnotetext[17]{Also at Dipartimento di Fisica, Universit\`a di Modena e Reggio Emilia, I-41125 Modena, Italy}
\footnotetext[18]{Also at Istituto di Fisica, Universit\`a di Urbino, I-61029 Urbino, Italy}
\footnotetext[19]{Present address: Helmholtz-Institut Mainz, Universit\"at Mainz, D-55099 Mainz, Germany}
\footnotetext[20]{Funded by the German Federal Minister for Education and Research under contract 05HK1UM1/1}
\footnotetext[21]{Present address: SLAC, Stanford University, Menlo Park, CA 94025, USA}
\footnotetext[22]{Present address: Dipartimento di Fisica dell'Universit\`a e Sezione dell'INFN di Padova, I-35131 Padova, Italy}
\footnotetext[23]{Present address: Instituto de F\'isica Corpuscular IFIC, Universitat de Valencia, E-46071 Valencia, Spain}
\footnotetext[24]{Present address: Dipartimento di Fisica dell'Universit\`a di Torino, I-10125 Torino, Italy}
\footnotetext[25]{Present address: Institut de Physique Nucl\'eaire de Lyon, Universit\'e Lyon I, F-69622 Villeurbanne, France}
\footnotetext[26]{Funded by the German Federal Minister for Research and Technology (BMBF) under contract 056SI74}
\footnotetext[27]{Present address: CIEMAT, E-28040 Madrid, Spain}
\footnotetext[28]{Funded by the Austrian Ministry for Traffic and Research under the contract GZ 616.360/2-IV GZ 616.363/2-VIII, and by the Fonds f\"ur Wissenschaft und Forschung FWF Nr.~P08929-PHY}

\newpage


\section*{Introduction}

Measurements of radiative non-leptonic kaon decays provide crucial tests of Chiral Perturbation Theory (ChPT) describing weak low energy processes. The
$K^\pm\to\pi^\pm\gamma\gamma$ decay (denoted $K_{\pi\gamma\gamma}$ below) has attracted the attention of theorists over the last 40 years~\cite{se72, ec88, da96, ge05}, but remains among the least experimentally studied kaon decays.

The standard kinematic variables for the $K_{\pi\gamma\gamma}$ decay are
\begin{displaymath}
z = \frac{(q_1+q_2)^2}{m_K^2} = \left(\frac{m_{\gamma\gamma}}{m_K}\right)^2, ~~~
y = \frac{p(q_1-q_2)}{m_K^2}~,
\end{displaymath}
where $q_{1,2}$ are the 4-momenta of the two photons, $p$ is the 4-momentum of the kaon, $m_{\gamma\gamma}$ is the di-photon invariant mass, and $m_K$ is the charged kaon mass. The physical region of the kinematic variables is~\cite{da96}
\begin{displaymath}
0\le z \le z_{\rm max}=(1-r_\pi)^2=0.515, ~~~~ 0\le y\le y_{\rm max}(z) = \frac{1}{2}\lambda^{1/2}\left(1, r_\pi^2,z\right),
\end{displaymath}
where $r_\pi=m_\pi/m_K$, $m_\pi$ is the charged pion mass and $\lambda(a,b,c)=a^2+b^2+c^2-2(ab+ac+bc)$.

The only published $K_{\pi\gamma\gamma}$ measurement to date comes from the BNL E787 experiment~\cite{ki97}: 31 $K^+$ decay candidates have been reported in the kinematic region $100~{\rm MeV}/c<p_\pi^*<180~{\rm MeV}/c$, where $p_\pi^*$ is the $\pi^+$ momentum in the $K^+$ rest frame (corresponding to $0.157<z<0.384$). A related decay mode
$K^\pm\to\pi^\pm\gamma e^+e^-$ (denoted $K_{\pi\gamma ee}$ below) has been measured from 120 decay candidates in the kinematic region $m_{\gamma ee}>260~{\rm MeV}/c^2$ or $z=(m_{\gamma ee}/m_K)^2>0.277$ by the NA48/2 experiment at CERN~\cite{ba08}.

A $K_{\pi\gamma\gamma}^\pm$ measurement with improved precision using minimum bias data sets collected by the NA48/2 experiment in 2003 and 2004 is reported here.

\section{Beam, detector and data samples}
\label{sec:experiment}

The NA48/2 experiment at CERN used simultaneous $K^+$ and $K^-$ beams produced by 400~GeV/$c$ primary SPS protons impinging on a beryllium target. Charged particles with momenta of $(60\pm3)$ GeV/$c$ were selected by an achromatic system of four dipole magnets, which split the two beams in the vertical plane and recombined them on a common axis. The beams then passed through collimators and a series of quadrupole magnets, and entered a 114~m long cylindrical vacuum tank with a diameter of 1.92 to 2.4~m containing the decay region. Both beams had a transverse size of about 1~cm, and were aligned with the longitudinal axis of the detector within 1~mm. The $K^+/K^-$ flux ratio was 1.79, and the fraction of beam kaons decaying in the vacuum tank was $22\%$.

The vacuum tank was followed by a magnetic spectrometer housed in a vessel filled with helium at nearly atmospheric pressure, separated from the vacuum by a thin ($0.3\%X_0$) $\rm{Kevlar}\textsuperscript{\textregistered}$ composite window. The quadrupole magnets mentioned earlier focused the beams to a waist near the centre of the spectrometer (the focusing was similar in the horizontal and vertical planes and for the $K^+$ and $K^-$ beams). An aluminium beam pipe of 158~mm outer diameter and 1.1~mm thickness traversing the centre of the spectrometer (and all the following detectors) allowed the undecayed beam particles to continue their path in vacuum. The spectrometer consisted of four drift chambers (DCH) with a transverse width of 2.9~m: DCH1, DCH2 located upstream and DCH3, DCH4 downstream of a dipole magnet that provided a horizontal transverse momentum kick of 120~MeV/$c$ for charged particles. Each DCH was composed of eight planes of sense wires and provided a space point resolution of $\sigma_x=\sigma_y=90~\mu$m. The spectrometer momentum resolution was $\sigma_p/p = (1.02 \oplus 0.044\cdot p)\%$, where $p$ is expressed in GeV/$c$. The spectrometer was followed by a plastic scintillator hodoscope (HOD) consisting of two planes with a transverse size of about 2.4~m, segmented in horizontal and vertical strips respectively, with each plane arranged in four quadrants. It provided trigger signals and time measurements of charged particles with a resolution of about 150~ps. The HOD was followed by a liquid krypton electromagnetic calorimeter (LKr), an almost homogeneous ionization chamber with an active volume of 7 m$^3$ of liquid krypton, $27X_0$ deep, segmented transversally into 13248 projective $\sim\!2\!\times\!2$~cm$^2$ cells and with no longitudinal segmentation. The LKr energy resolution was $\sigma_E/E=(3.2/\sqrt{E}\oplus9/E\oplus0.42)\%$, and its spatial resolution for the transverse coordinates $x$ and $y$ of an isolated electromagnetic shower was $\sigma_x=\sigma_y=(4.2/\sqrt{E}\oplus0.6)$~mm, where $E$ is expressed in GeV. The LKr was followed by a hadronic calorimeter and a muon detector, both not used in the present analysis. A detailed description of the detector can be found in Ref.~\cite{fa07}.

The experiment collected data during two high intensity runs in 2003 and 2004 (with about $3\times 10^6$ $K^\pm$ entering the decay volume per SPS spill of 4.8~s duration), in about 100 days of efficient data taking in total. A multi-level trigger was employed to collect $K^\pm$ decays with at least three charged tracks in the final state, as well as $K^\pm\to\pi^\pm\pi^0\pi^0$ decays~\cite{ba07}: it had low efficiency for the $K_{\pi\gamma\gamma}$ decays, potentially leading to sizeable systematic uncertainties. Therefore the present $K_{\pi\gamma\gamma}$ measurement is based on two special $K^\pm$ decay samples collected at $\sim 10\%$ the nominal beam intensity during 12 hours in 2003 and 54 hours in 2004 with a minimum bias trigger condition: a time coincidence of signals in both HOD planes within the same quadrant and an energy deposit of at least 10~GeV in the LKr calorimeter.

A GEANT3-based~\cite{geant} Monte Carlo (MC) simulation including kaon beam line, detector geometry and material description is used to evaluate the detector response.

\section{Data analysis}

\subsection{Measurement method}
\label{sec:method}

The $K_{\pi\gamma\gamma}$ decay rate is measured with respect to the normalization decay chain with a large and well known branching fraction~\cite{pdg}: the $K^\pm\to\pi^\pm\pi^0$ decay (denoted $K_{2\pi}$ below) followed by the $\pi^0\to\gamma\gamma$ decay (denoted $\pi^0_{\gamma\gamma}$ below). Signal and normalization samples have been collected with the same trigger logic. With this approach, the branching ratio of $K_{\pi\gamma\gamma}$ decay can be computed as
\begin{displaymath}
{\cal B}(K_{\pi\gamma\gamma}) = \frac{N_{\pi\gamma\gamma}^\prime}{N_{2\pi}^\prime} \cdot \frac{A_{2\pi}}{A_{\pi\gamma\gamma}} \cdot \frac{\varepsilon_{2\pi}}{\varepsilon_{\pi\gamma\gamma}} \cdot {\cal B}(K_{2\pi}){\cal B}(\pi^0_{\gamma\gamma}),
\end{displaymath}
where $N_{\pi\gamma\gamma}^\prime$ and $N_{2\pi}^\prime$ are the numbers of reconstructed signal and normalization events (with backgrounds subtracted), $A_{\pi\gamma\gamma}$ and $A_{2\pi}$ are the acceptances of the signal and normalization selections, and $\varepsilon_{\pi\gamma\gamma}$ and $\varepsilon_{2\pi}$ are the corresponding trigger efficiencies.

The acceptances are computed with MC simulations. However the signal acceptance $A_{\pi\gamma\gamma}$ is not uniform over the kinematical space, and therefore depends in general on the assumed kinematic distribution. Trigger efficiencies have been measured in dedicated data studies and found to have similar values for the signal, normalization and background decay modes with similar final state topologies. Therefore they cancel to first order both while correcting the ratio of signal to normalization counts and while subtracting background from the signal counts. The residual systematic effects are well below the statistical precision of the measurement, as detailed in Section~\ref{sec:syst}.


\subsection{Event reconstruction and selection}
\label{sec:selection}

Trajectories and momenta of charged particles are reconstructed from hits and drift times in the spectrometer using a detailed magnetic field map. Fine calibrations of the spectrometer field integral and DCH alignment are based on measurements of the mean reconstructed $K^\pm\to3\pi^\pm$ invariant mass. Clusters of energy deposition in the LKr calorimeter are found by locating the maxima in the digitized pulses from individual cells in space and time. Cluster positions are estimated using the centres of gravity of the energy deposition in $3\times3$ cells, while their energies are estimated as sums of energies deposited in the cells within 11~cm from the maxima. The reconstructed energies are corrected for energy deposited outside the cluster boundary, energy sharing and losses in inactive cells (0.8\% of the total number). Further details about the reconstruction procedure can be found in Ref.~\cite{fa07}.

The signal ($K_{\pi\gamma\gamma}$) and normalization ($K_{2\pi}$, $\pi^0_{\gamma\gamma}$) decay modes are characterized by the same set of particles in the final state. Therefore the following principal selection criteria are common for the two modes, leading to cancellation of systematic effects.
\begin{itemize}
\item Exactly one reconstructed charged particle track ($\pi^\pm$ candidate) geometrically consistent with originating from a $K^\pm$ decay is required.
    The geometrical consistency is determined by reconstructing the decay vertex as the point of closest approach of the track (extrapolated from the spectrometer
    upstream into the vacuum tank) and the detector axis, taking into account the stray magnetic field. The reconstructed closest distance of approach (CDA) of the track to the detector axis is required to be less than 3.5~cm. The width (rms) of the CDA distribution for $K_{2\pi}$ events without $\pi^\pm$ decays in flight is 0.5~cm, dominated by the beam transverse size. The reconstructed kaon decay vertex should be located within a 98~m long fiducial volume in the upstream part of the vacuum tank.
\item Track impact points in the DCH, HOD and LKr calorimeter front planes should be within the corresponding fiducial acceptances, including appropriate separations from detector edges and inactive LKr cells.
\item The reconstructed track momentum should be between 10 and 40 GeV/$c$. The lower cut results in a relative $K_{\pi\gamma\gamma}$ acceptance loss of about 10\% (assuming a ChPT kinematic distribution), reducing the $K^\pm\to\pi^\pm\pi^0\pi^0$ background by about 40\%. The upper cut, resulting in no $K_{\pi\gamma\gamma}$ acceptance loss and decreasing the $K_{2\pi}$ acceptance by about 5\% relative, is equivalent to a lower limit on the total energy of the two photons and ensures the high efficiency of the LKr trigger condition.
\item The charged pion ($\pi^\pm$) is identified by the ratio $E/p$ of energy deposition in the LKr calorimeter to momentum measured by the spectrometer: $E/p<0.85$. This decreases electron contamination in the pion sample by about two orders of magnitude and reduces the backgrounds from kaon decays to electrons such as $K^\pm\to\pi^0e^\pm\nu(\gamma)$ to a negligible level. The $\pi^\pm$ identification efficiency, discussed in Section~\ref{sec:syst}, is about 98.5\%.
\item LKr energy deposition clusters in time with the track ($\pm15$~ns) and separated by at least 25~cm from the track impact point are considered as photon candidates. The presence of exactly two photon candidates is required. The candidates should be within the fiducial LKr acceptance and separated from inactive LKr cells. The distance between the two candidates should be larger than 20~cm, and their energies should exceed 3 GeV. The latter two requirements do not lead to $K_{\pi\gamma\gamma}$ acceptance loss (due to the $m_{\gamma\gamma}$ cut discussed below) but reduce the $K_{2\pi}$ acceptance by about 6\% relative.
\item To suppress backgrounds due to LKr cluster merging, an energy-dependent upper cut on the LKr cluster transverse width is applied to the photon candidates. The criterion has been established by analyzing the width distributions of isolated electromagnetic clusters separately for data and MC simulated events. It reduces background in the $K_{\pi\gamma\gamma}$ sample by about a factor of 2 (as discussed in Section~\ref{sec:bkg}) with a 0.7\% relative acceptance reduction for both $K_{\pi\gamma\gamma}$ and $K_{2\pi}$ decays.
\item Photon trajectories and 4-momenta are reconstructed assuming that the photons originate from the decay vertex defined above. The trajectories are required not to intersect the beam pipe and inner DCH flanges to avoid energy and momentum mismeasurement due to showering in the material: at least 11~cm separation from the detector axis in DCH transverse planes is required.
\item The reconstructed total $\pi^\pm\gamma\gamma$ momentum should be between 55 and 65~GeV/$c$ and the transverse momentum with respect to the detector axis should be $p_T^2<0.5\times 10^{-3}~({\rm GeV}/c)^2$, which is consistent with the beam momentum spectrum. The relative acceptance losses due to these conditions are below 1\% for both $K_{\pi\gamma\gamma}$ and $K_{2\pi}$ decays.
\item The reconstructed $\pi^\pm\gamma\gamma$ ($\pi^\pm\pi^0$) invariant mass should be between 0.48 and 0.51~GeV/$c^2$. The corresponding mass resolutions are 5.9 (3.9)~MeV/$c^2$ for the $K_{\pi\gamma\gamma}$ ($K_{2\pi}$) decays.
\end{itemize}
The $K_{\pi\gamma\gamma}$ and $K_{2\pi}$ selection conditions differ only in the di-photon invariant mass requirement.
\begin{itemize}
\item For $K_{\pi\gamma\gamma}$, the signal kinematic region is defined as $z>0.2$. The low $z$ region is dominated by the $K_{2\pi}$ background and other backgrounds from $\pi^0_{\gamma\gamma}$ decays peaking at $z=(m_{\pi^0}/m_K)^2=0.075$. Earlier analyses of $K_{\pi\gamma\gamma}$ and $K_{\pi\gamma ee}$ decays~\cite{ki97,ba08} are also restricted to kinematic regions above the $\pi^0$ peak for the same reason. As discussed in Section~\ref{sec:fits}, the expected $K_{\pi\gamma\gamma}$ acceptance loss assuming a ChPT kinematic distribution is only a few percent. The resolution on the $z$ variable increases from $\delta z=0.005$ at $z=0.2$ to $\delta z=0.03$ at $z_{\rm max}=0.515$.
\item For $K_{2\pi}$, the reconstructed di-photon mass should be consistent with the nominal $\pi^0$ mass~\cite{pdg}: $|m_{\gamma\gamma}-m_{\pi^0}|<10~{\rm MeV}/c^2$, equivalent to $0.064<z<0.086$. The resolution on the $\pi^0$ mass is $\delta m_{\gamma\gamma}=1.6~{\rm MeV}/c^2$, or $\delta z=0.002$.
\end{itemize}

\begin{figure}[t]
\begin{center}
\resizebox{0.50\textwidth}{!}{\includegraphics{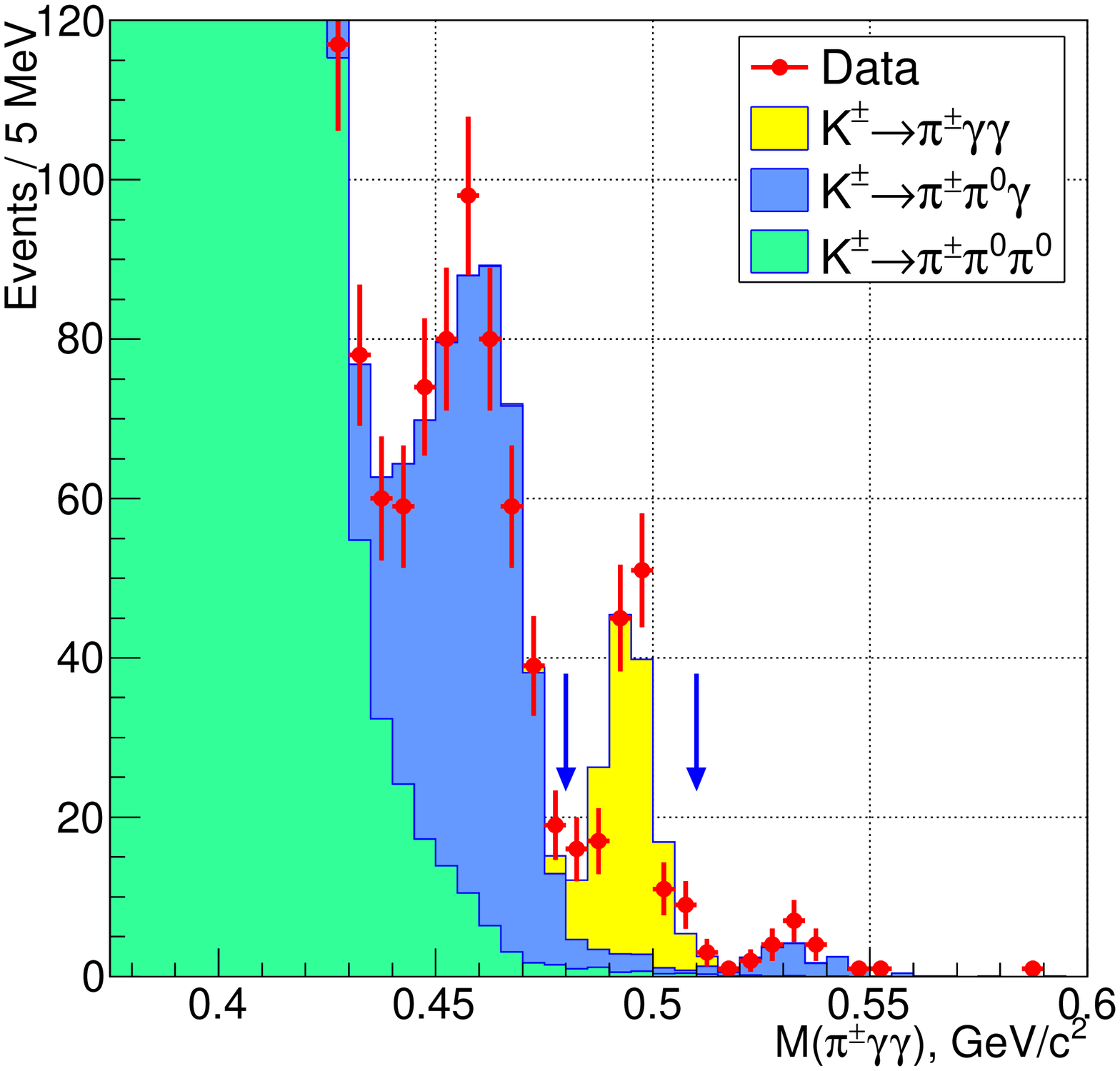}}%
\resizebox{0.50\textwidth}{!}{\includegraphics{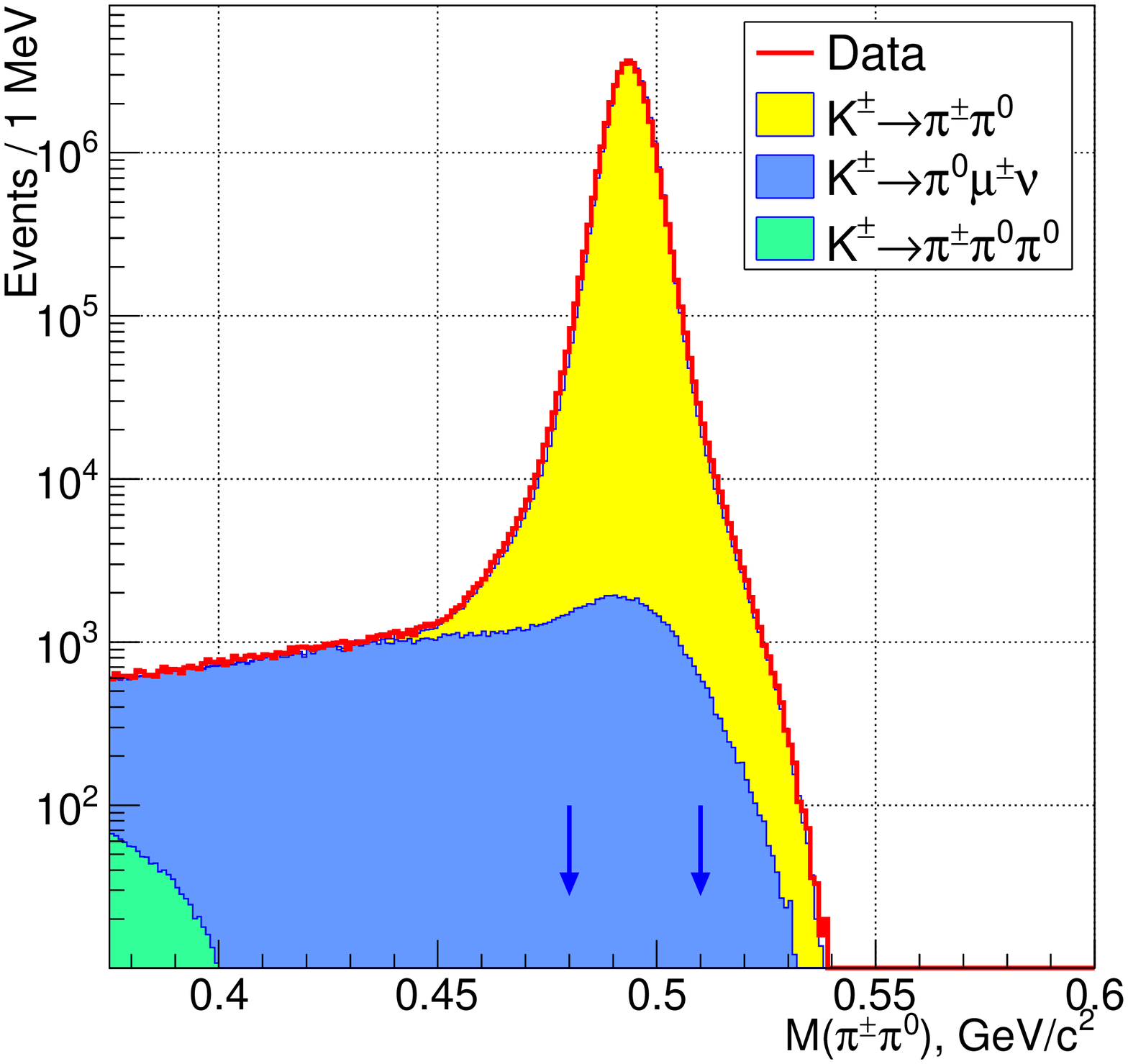}}
\put(-428,201){\bf\large (a)} \put(-201,201){\bf\large (b)}
\end{center}
\vspace{-16mm} \caption{Invariant mass distributions of (a) $\pi^\pm\gamma\gamma$ and (b) $\pi^\pm\pi^0$ compared with the sums of estimated signal and background components.
The estimated $K_{\pi\gamma\gamma}$ signal corresponds to the result of a ChPT ${\cal O}(p^6)$ fit. The limits of the signal regions are indicated with vertical arrows.}
\label{fig:mass}
\end{figure}

\begin{figure}[t]
\begin{center}
\resizebox{0.50\textwidth}{!}{\includegraphics{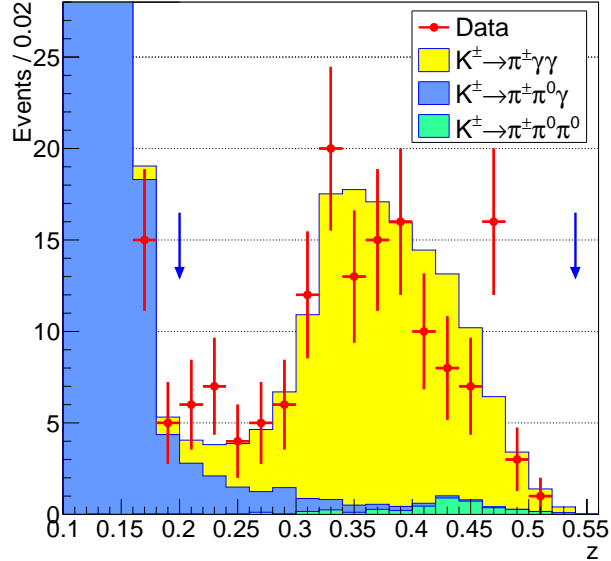}}
\end{center}
\vspace{-16mm} \caption{Reconstructed $z=(m_{\gamma\gamma}/m_K)^2$ distribution of the $K_{\pi\gamma\gamma}$ candidates and estimated signal and background contributions. The estimated signal corresponds to the result of a ChPT ${\cal O}(p^6)$ fit. The limits of the signal region are indicated with vertical arrows.}
\label{fig:z}
\end{figure}

The $\pi^\pm\gamma\gamma$ and $\pi^\pm\pi^0$ invariant mass spectra of the selected signal and normalization candidates are displayed in Fig.~\ref{fig:mass} together with the expectations for the signal and background contributions evaluated with MC simulations. The number of reconstructed $K_{\pi\gamma\gamma}$ candidates is $N_{\pi\gamma\gamma}=149$, of which 97 (52) are $K^+$ ($K^-$) decay candidates. The number of reconstructed $K_{2\pi}$ candidates is $N_{2\pi} = 3.628\times 10^7$, of which $2.321~(1.307)\times 10^7$ are $K^+$ ($K^-$) decay candidates. The ratios of
the numbers of $K_{\pi\gamma\gamma}$/$K_{2\pi}$ candidates are consistent for $K^+$ and $K^-$ decays, as the NA48/2 geometrical acceptance is highly charge-symmetric by design~\cite{ba07}. The reconstructed $z$ spectrum of the $K_{\pi\gamma\gamma}$ candidates is presented in Fig.~\ref{fig:z}.

\subsection{Backgrounds}
\label{sec:bkg}

The only significant background to the normalization mode ($K_{2\pi}$, $\pi^0_{\gamma\gamma}$) comes from the $K^\pm\to\pi^0\mu^\pm\nu$ decay (denoted $K_{\mu3}$ below) followed by $\pi^0_{\gamma\gamma}$. The relative background contamination is estimated to be $R = {\cal B}(K_{\mu3})A(K_{\mu3}) / {\cal B}(K_{2\pi})A(K_{2\pi})= 0.13\%$, where ${\cal B}$ denote the nominal branching fractions~\cite{pdg}, and $A(K_{2\pi})=19.18\%$, $A(K_{\mu3})=0.15\%$ are the acceptances of the $K_{2\pi}$ selection for $K_{2\pi}$ and $K_{\mu3}$ decays followed by $\pi^0_{\gamma\gamma}$ decays evaluated with MC simulation. The product of the number of $K^\pm$ decays in the fiducial volume and the trigger efficiency for the $K_{2\pi}$ sample is computed from the number of reconstructed normalization candidates $N_{2\pi}$ as
\begin{displaymath}
N_K = \frac{N_{2\pi}}{{\cal B}(K_{2\pi}){\cal B} (\pi^0_{\gamma\gamma})A(K_{2\pi})(1+R)}=(0.925\pm0.004)\times 10^9,
\end{displaymath}
where the uncertainty is due to the limited precision on the external input ${\cal B}(K_{2\pi})$. The number of background events in the $K_{\pi\gamma\gamma}$ sample is evaluated as
\begin{displaymath}
N^B = N_K \times \sum_i {\cal B}^B_i A^B_i,
\end{displaymath}
where the sum runs over the background kaon decay modes, ${\cal B}^B_i$ are the
corresponding branching ratios and $A^B_i$ are their geometrical acceptances within the $K_{\pi\gamma\gamma}$ selection evaluated with MC simulation. As discussed in Section~\ref{sec:method}, this approach relies on the cancellation of the trigger efficiencies.

The dominant background to the $K_{\pi\gamma\gamma}$ decay comes from the $K^\pm\to\pi^\pm\pi^0\gamma$ inner brems\-strah\-lung (IB) decay, simulated according to Ref.~\cite{ga06}, followed by $\pi^0_{\gamma\gamma}$.
Two contributions to the IB process have been considered separately to improve the statistical precision: a) the component with the radiative photon energy in the kaon rest frame $E_\gamma^*>10~{\rm MeV}$, accounting for 0.32\% of the decay rate and about 90\% of the background; b) the remaining component with $E_\gamma^*\le 10~{\rm MeV}$. The smaller contributions from the $K^\pm\to\pi^\pm\pi^0\gamma$ direct emission (DE) and interference (INT) terms followed by $\pi^0_{\gamma\gamma}$ decay are simulated using the expected ChPT phase space distributions~\cite{ch67,dafne} and the measured decay rates~\cite{ba10}: the corresponding partial decay rates integrated over the phase space are ${\cal B}_{\rm DE}=(6.9\pm0.4)\times 10^{-6}$ and ${\cal B}_{\rm INT}=(-6.0\pm1.3)\times 10^{-6}$. The $K^\pm\to\pi^\pm\pi^0\gamma$ (IB, DE, INT) decays can produce a $K_{\pi\gamma\gamma}$ signature by the following mechanisms.
\begin{itemize}
\item In the mass region $m_{\pi\gamma\gamma}<0.48~{\rm GeV}/c^2$: a photon from the $\pi^0_{\gamma\gamma}$ decay is outside the LKr acceptance. This contribution comes mainly from the high-$E^*_\gamma$ IB component as the selection requires the radiative photon to produce an LKr cluster with an energy of at least 3~GeV.
\item In the signal $m_{\pi\gamma\gamma}$ region: LKr clusters produced by the radiative photon and a photon from the $\pi^0_{\gamma\gamma}$ decay merge, resulting in the reconstructed $z$ variable above $z=(m_{\pi^0}/m_K)^2=0.075$. This irreducible background comes mainly from the high-$E^*_\gamma$ IB component.
\item In the mass region $m_{\pi\gamma\gamma}>0.51~{\rm GeV}/c^2$: the radiative photon is undetected, while a photon from the $\pi^0_{\gamma\gamma}$ decay converts in the spectrometer between DCH1 and the magnet ($\gamma\to e^+e^-$), resulting in two LKr clusters but no reconstructed tracks.\footnote{Track reconstruction requires space points in each of DCH1, DCH2 and DCH4.} The other photon from the $\pi^0_{\gamma\gamma}$ decay either forms a merged LKr cluster or is outside the LKr acceptance. This contribution comes mainly from the low-$E^*_\gamma$ IB component.
\end{itemize}
The latter two components involve merging of LKr electromagnetic clusters and are reduced by the cluster transverse width cut (see Section~\ref{sec:selection}). The total $K^\pm\to\pi^\pm\pi^0\gamma$ background is estimated to be $11.4\pm0.6$ events, where the uncertainty comes from MC simulation statistics.

Another source of background to the $K_{\pi\gamma\gamma}$ decay is the $K^\pm\to\pi^\pm\pi^0\pi^0$ decay followed by $\pi^0_{\gamma\gamma}$ decays.
It enters the signal region predominantly due to the two photons out of the four (coming from decays of different $\pi^0$ to satisfy the $z>0.2$ requirement for the remaining pair) missing the LKr fiducial area. There is also a component with one or two pairs of photons (from decays of different $\pi^0$) producing merged LKr clusters. The invariant mass of the four photons corresponds to $z\ge(2m_{\pi^0}/m_K)^2 = 0.299$. The $z$ variable reconstructed with the two clusters also satisfies this condition (Fig.~\ref{fig:z}) because the lost photons (if any) are soft, as imposed by the total momentum and invariant mass requirements. The background is estimated to be $4.1\pm0.4$ events, where the uncertainty comes from MC simulation statistics.

The total estimated background in the $K_{\pi\gamma\gamma}$ sample amounts to $15.5\pm0.7$ events, where the error is MC statistical. The data/MC agreement of the distributions outside the signal regions (Fig.~\ref{fig:mass}, \ref{fig:z}) validates the background estimates to a good accuracy.


\subsection{Model-independent rate measurement}
\label{sec:model-independent}

Partial $K_{\pi\gamma\gamma}$ branching fractions ${\cal B}_j$ in  8 bins of the $z$ variable defined in Table~\ref{tab:brmi} are evaluated as
\begin{displaymath}
{\cal B}_j = (N_j - N^B_j) / (N_K A_j),
\end{displaymath}
where $N_j$ is the number of reconstructed $K_{\pi\gamma\gamma}$ candidates, $N^B_j$ is the number of background events and $A_j$ is the signal acceptance in bin $j$ (the latter two quantities are estimated from MC simulation). Background evaluation in bins of $z$ is similar to that in the total $K_{\pi\gamma\gamma}$ sample described in Section~\ref{sec:bkg}. Trigger efficiency cancels at this stage, as discussed in Section~\ref{sec:method}. The resulting measurement of the $z$ spectrum is model-independent because the considered $z$ bin width is sufficiently small for the acceptances $A_j$ to have a negligible dependence on the assumed $K_{\pi\gamma\gamma}$ kinematical distribution. In addition, the $y$-dependence of the differential decay rate expected within the ChPT framework~\cite{da96,ge05} is weak (see Section~\ref{sec:fits}), and the $y$-dependence of acceptance is also weak. The values of $N_j$, $N^B_j$ and $A_j$ and the evaluated ${\cal B}_j$ with their statistical uncertainties are presented in Table~\ref{tab:brmi}. The model-independent branching fraction in the kinematic region $z>0.2$ is computed by summing over the $z$ bins:
\begin{displaymath}
{\cal B}_{\rm MI}(z>0.2) = \sum\limits_{j=1}^{8}{\cal B}_{j} = (0.877 \pm 0.087_{\rm stat}) \times 10^{-6}.
\end{displaymath}

\begin{table}[tb]
\begin{center}
\caption{Numbers of signal and background events $N_j$ and $N^B_j$, signal acceptances $A_j$ and model-independent branching ratios ${\cal B}_j$ evaluated in $z$ bins. The quoted uncertainties are statistical. Signal acceptance reduces to zero at the endpoint $z_{\rm max}$, as the $\pi^\pm$ at rest in the $K^\pm$ centre of mass frame propagates in the beam pipe. The acceptance for the normalization mode is $A_{2\pi}=0.1918$, as reported in Section~\ref{sec:bkg}.}
\vspace{1mm}
\label{tab:brmi}
\begin{tabular}{crrcc}
\hline
$z$ range & $N_j$ & $N^B_j$ & $A_j$ & ${\cal B}_j\times 10^6$\\
\hline
0.20--0.24 & 13 & 4.89 & 0.194 & $0.045 \pm 0.020$\\
0.24--0.28 &  9 & 2.73 & 0.198 & $0.034 \pm 0.016$\\
0.28--0.32 & 18 & 2.33 & 0.194 & $0.087 \pm 0.024$\\
0.32--0.36 & 33 & 1.30 & 0.190 & $0.180 \pm 0.033$\\
0.36--0.40 & 31 & 0.98 & 0.184 & $0.177 \pm 0.033$\\
0.40--0.44 & 18 & 1.61 & 0.173 & $0.103 \pm 0.027$\\
0.44--0.48 & 23 & 1.21 & 0.135 & $0.175 \pm 0.038$\\
$z>0.48$   &  4 & 0.52 & 0.049 & $0.076 \pm 0.044$\\
\hline
\end{tabular}
\end{center}
\vspace{-10mm}
\end{table}

\subsection{Measurement of ChPT parameters}
\label{sec:fits}

Given the limited size of the data sample, the ChPT formulation of Ref.~\cite{da96}, which involves fewer free parameters than a similar formulation of Ref.~\cite{ge05}, is considered in this analysis. The $K_{\pi\gamma\gamma}$ decay receives no tree-level ${\cal O}(p^2)$ contribution, and the differential decay rate for leading order ${\cal O}(p^4)$ and including next-to-leading order ${\cal O}(p^6)$ contributions can be parameterized as follows:
\begin{displaymath}
\frac{\partial\Gamma}{\partial y \partial z}(\hat c, y, z) = \frac{m_K}{2^9\pi^3}\left[z^2\left(|A(\hat c, z, y^2) + B(z)|^2 + |C(z)|^2 \right) + \left(y^2-\frac{1}{4}\lambda(1,r_\pi^2,z)\right)^2\left|B(z)\right|^2\right].
\end{displaymath}
Here $A(\hat c, z, y^2)$ and $B(z)$ are loop amplitudes (the latter appears at next-to-leading order and dominates the differential rate at low $z$), and $C(z)$ is a pole amplitude contributing a few percent to the total decay rate. The rate and spectrum are determined by a single ${\cal O}(1)$ parameter $\hat c$ whose value is a priori unknown. An additional loop amplitude $D$ entering the complete formulation vanishes at ${\cal O}(p^6)$ for the $K_{\pi\gamma\gamma}$ process~\cite{da96}, though it does contribute at this order to the $K_{\pi\gamma ee}$ process with an off-shell photon~\cite{ga99}. The $y^2$-dependence of the differential decay rate arises only at ${\cal O}(p^6)$ and is weak: e.g. for $\hat c=2$, the relative variation of $\partial\Gamma/\partial z\partial y$ over $y$ for a fixed $z$ is below 14\% for $z>0.2$ and below 6\% for $z>0.25$. The explicit expressions for the above amplitudes are given in Ref.~\cite{da96}.

The ChPT description involves a number of external inputs. The $G_8$ parameter entering both ${\cal O}(p^4)$ and ${\cal O}(p^6)$ descriptions is fixed in this analysis according to Ref.~\cite{ci12}. The ${\cal O}(p^6)$ framework additionally involves 7 parameters of the $K_{3\pi}$ decay amplitude fixed in this analysis to those fitted to the experimental data~\cite{bi03}, and 3 polynomial contributions $\eta_i$ ($i=1;2;3$) fixed to $\eta_i=0$. The parameter $\hat c$ enters the ${\cal O}(p^6)$ differential decay rate via a linear combination $\hat c^* = \hat c - 2(m_\pi/m_K)^2\eta_1 - 2\eta_2 - 2\eta_3$. Therefore setting $\eta_i=0$ is equivalent to measuring $\hat c^*$, and $\hat c$ can be computed for any assumed values of $\eta_i$.

\begin{figure}[tb]
\begin{center}
\resizebox{0.50\textwidth}{!}{\includegraphics{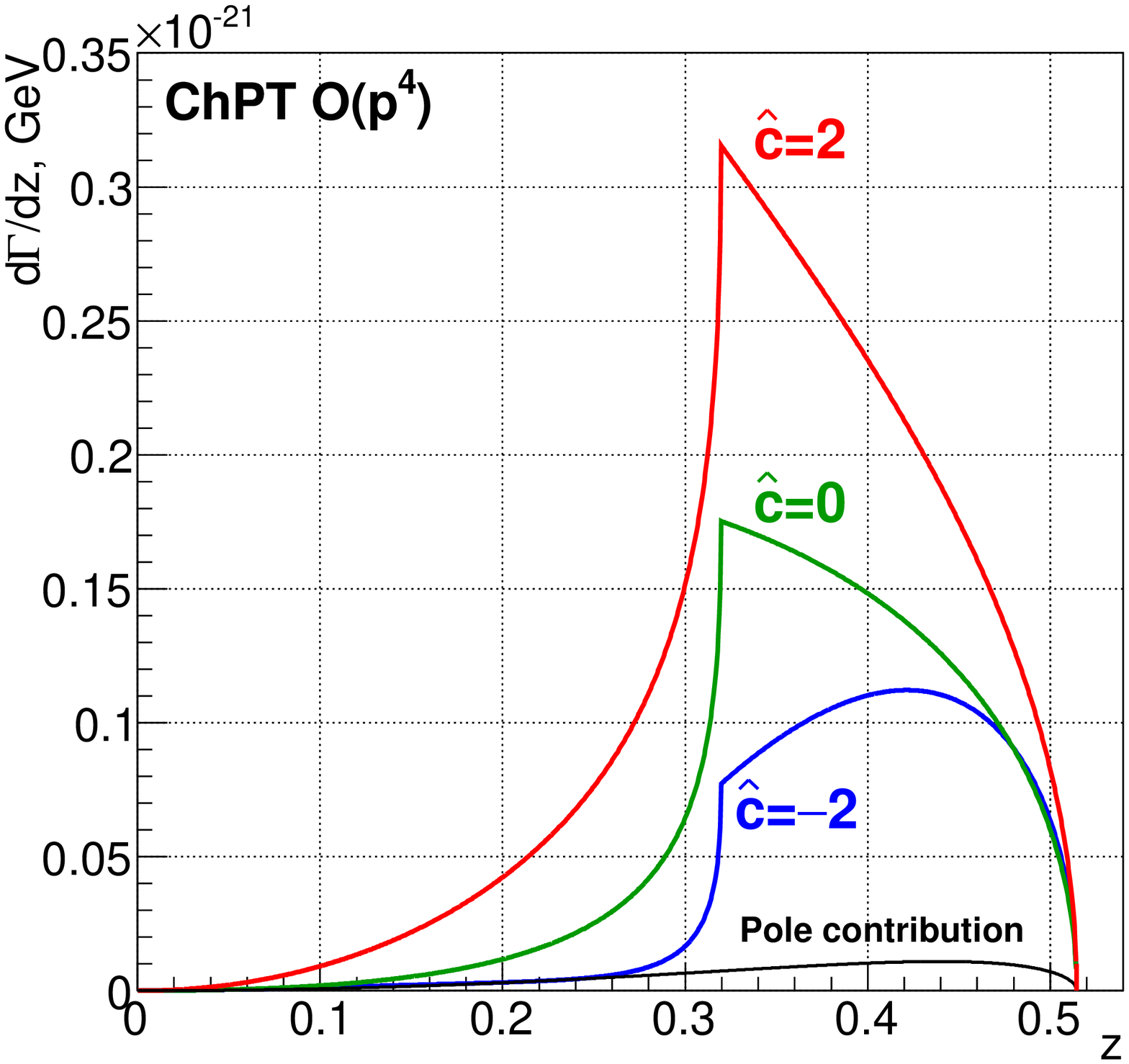}}%
\resizebox{0.50\textwidth}{!}{\includegraphics{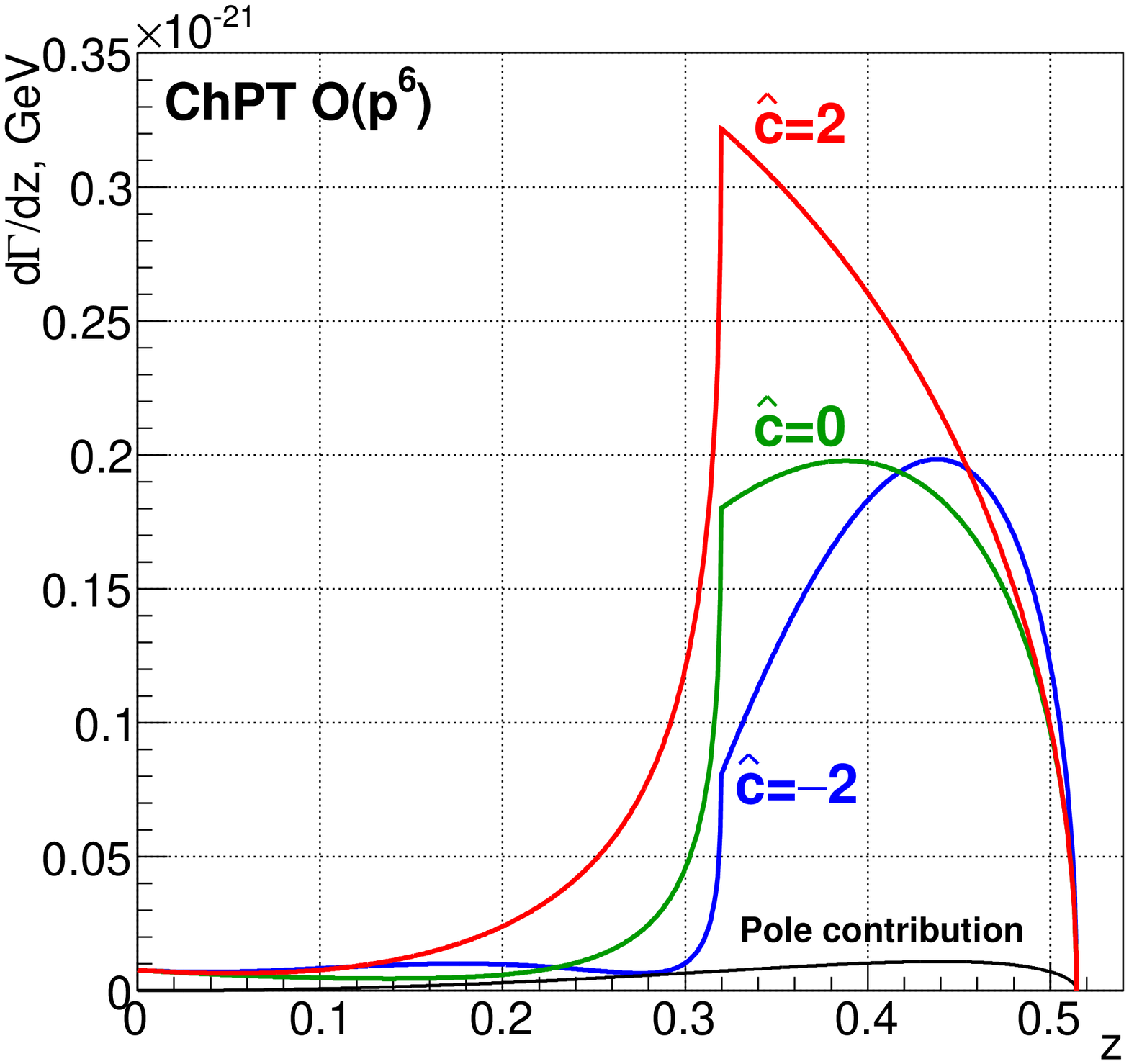}}
\end{center}
\vspace{-16mm} \caption{Differential rate $d\Gamma/dz$ according to the ${\cal O}(p^4)$ and ${\cal O}(p^6)$ descriptions~\cite{da96} for several values of $\hat c$. The $\hat c$-independent pole contribution from the $C(z)$ amplitude is also shown. The external parameters are fixed as indicated in Table~\ref{tab:extpar}.}
\label{fig:chpt}
\end{figure}

The considered values of the external parameters are listed in Table~\ref{tab:extpar}.
The corresponding ChPT ${\cal O}(p^4)$ and ${\cal O}(p^6)$ predictions~\cite{da96} for the differential decay rate are illustrated in Fig.~\ref{fig:chpt}. Their main features are: a) a cusp at the di-pion threshold $z_{\rm th}=4r_\pi^2=0.320$ generated by the pion loop amplitude; b) non-zero differential rate at $z=0$ generated by the $B(z)$ amplitude at next-to-leading order ${\cal O}(p^6)$. The branching ratio is expected to be ${\cal B}(K_{\pi\gamma\gamma}) \sim 10^{-6}$.

\begin{table}[tb]
\begin{center}
\caption{Values of the external parameters considered in this analysis. The notation is introduced in Ref.~\cite{da96,ci12,bi03}.}
\vspace{1mm}
\label{tab:extpar}
\begin{tabular}{cr|cr|cr}
\hline
Parameter & Value & Parameter & Value & Parameter & Value\\
\hline
$G_8m_K^2\times 10^6$ &  $2.202$ & $\beta_1\times 10^8$  & $-27.06$ & $\zeta_1\times 10^8$  &  $-0.40$\\
$\alpha_1\times 10^8$ &  $93.16$ & $\beta_3\times 10^8$  &  $-2.22$ & $\xi_1\times 10^8$    &  $-1.83$\\
$\alpha_3\times 10^8$ &  $-6.72$ & $\gamma_3\times 10^8$ &   $2.95$ &$\eta_i$              &      $0$\\
\hline
\end{tabular}
\end{center}
\vspace{-13mm}
\end{table}

To measure the values of the $\hat c$ parameter in the ChPT ${\cal O}(p^4)$ and ${\cal O}(p^6)$ frameworks, fits to the reconstructed $z$ spectrum (Fig.~\ref{fig:z}) have been performed by maximizing the log-likelihood
\begin{displaymath}
\ln{\cal L} = \sum\limits_{i=1}^{17} \left[ n_i \ln m_i - m_i - \ln(n_i!)\right].
\end{displaymath}
The sum runs over bins of the reconstructed $z$ variable in the range $0.2<z<0.54$ (bin width is $\delta z=0.02$), $n_i$ are the numbers of observed data events in the bins, and $m_i(\hat c) = m_i^{S}(\hat c) + m_i^{B}$ are the expected numbers of events for a given value of $\hat c$, including signal and background components $m_i^{S}(\hat c)$ and $m_i^{B}$. The quantities $m_i(\hat c)$ are computed using the number of $K^\pm$ decays in the fiducial volume $N_K$ measured from the normalization sample (Section~\ref{sec:bkg}), the expected ChPT $K_{\pi\gamma\gamma}$ differential decay rate for a given $\hat c$ value~\cite{da96}, and the acceptances of the $K_{\pi\gamma\gamma}$ selection for signal and backgrounds evaluated from MC simulations. The highest bin is above the $K_{\pi\gamma\gamma}$ kinematic endpoint and is populated due to resolution effects (see Fig.~\ref{fig:z}). The results of the fits to the ${\cal O}(p^4)$ and ${\cal O}(p^6)$ formulations~\cite{da96} are
\begin{displaymath}
\hat c_4 = 1.37\pm 0.33_{\rm stat}, ~~~ \hat c_6 = 1.41\pm 0.38_{\rm stat}.
\end{displaymath}
A binned Kolmogorov--Smirnov test~\cite{james} for the ChPT ${\cal O}(p^4)$ and ${\cal O}(p^6)$ descriptions yields $p$-values of 46\% and 59\%, respectively: the data are consistent with both considered descriptions. The total decay rate has an approximately parabolic dependence on $\hat c$~\cite{da96}; the corresponding local maximum of the likelihood function at $\hat c_4 \approx -8$, $\hat c_6 \approx -6$ is ruled out by the data. The $z$ spectrum corresponding to the ${\cal O}(p^6)$ fit shown in Fig.~\ref{fig:z} supports the ChPT prediction of a cusp at the di-pion threshold.

\subsection{Systematic effects}
\label{sec:syst}

The largest systematic uncertainty comes from the background estimate in the $K_{\pi\gamma\gamma}$ sample. As discussed in Section~\ref{sec:bkg}, the background comes mainly from $K^\pm$ decays with a pair of nearby LKr clusters produced by two photons reconstructed as a single cluster. Therefore the background estimation relies on the simulation of LKr cluster merging. To quantify the systematic effect, stability of the results with respect to the variation of the LKr cluster transverse width cut has been studied. The variation includes the removal of the cut, leading to a background enhancement by a factor of $\sim 2$, which is largely compensated by a similar increase in the background estimate. In another check, artificial merging of pairs of nearby reconstructed clusters has been introduced for both data and MC simulated samples, with pairs of clusters separated by less than a certain {\it merging distance} replaced by a single merged cluster. A stability test has been performed with respect to the variation of the merging distance parameter from zero (the standard selection) to 6.5~cm (at which distance clusters are normally resolved), leading to background enhancement by a factor of $\sim 2.5$. These tests have not revealed any systematic effects within their statistical sensitivity. Maximum variations of the results are conservatively considered as the systematic uncertainties due to background estimation: $\delta {\cal B}_{\rm MI}(z>0.2)=0.017\times 10^{-6}$, $\delta\hat c_4=0.14$, $\delta\hat c_6=0.11$. The uncertainties due to the MC statistical errors of background estimates are negligible with respect to the systematic uncertainties quoted above.


The HOD trigger efficiency for 1-track events has been measured to be 99.75\% and geometrically uniform using control triggers requiring activity in the LKr~\cite{ba07}. The upper track momentum (40~GeV/$c$) and lower total momentum (55~GeV/$c$) selection conditions constrain the LKr energy deposit to be above 15 GeV, which is higher than the 10 GeV trigger threshold. The corresponding LKr trigger efficiency has been measured to be above 99\% using a HOD control trigger. Efficiencies of both HOD and LKr trigger conditions largely cancel between the signal, normalization and background channels for the adopted event selection due to the absence of significant geometric or energy dependences. The residual systematic effect is negligible.

The $\pi^\pm$ identification efficiency due to the $E/p<0.85$ condition (Section~\ref{sec:selection}) is not perfectly reproduced by the MC simulation, due to the limited precision of hadronic shower description. It has been measured from samples of $K_{2\pi}$ and $K^\pm\to3\pi^\pm$ decays to vary from 98.6\% at $p=10~{\rm GeV}/c$ to 98.3\% at $p=40~{\rm GeV}/c$. It largely cancels between the signal, normalization and background channels separately for data and MC simulated samples due to its geometric uniformity and weak momentum dependence. The residual systematic bias is significantly below the statistical uncertainties.

The uncertainties due to the limited accuracy of geometrical acceptance evaluation are well below the statistical precision. The systematic effects due to accidental activity are negligible, as the data sample was collected at low beam intensity.

The uncertainty on the total number of kaon decays in the fiducial volume due to the limited precision on the external input ${\cal B}(K_{2\pi})$ is $\delta N_K/N_K=0.4\%$. It translates into negligible uncertainties on the results: $\delta {\cal B}_{\rm MI}(z>0.2) = 0.004\times 10^{-6}$, $\delta\hat c_4 = \delta\hat c_6 = 0.01$.


\section{Results and conclusions}

A sample of 149 $K^\pm_{\pi\gamma\gamma}$ decay candidates with an estimated background contamination of $15.5\pm0.7$ events collected by the NA48/2 experiment at CERN with minimum bias trigger conditions in 2003 and 2004 has been analyzed. Using the $K_{2\pi}$ decay followed by $\pi^0_{\gamma\gamma}$ as normalization mode, the model-independent (MI) $K^\pm_{\pi\gamma\gamma}$ branching ratio in the kinematic region $z>0.2$ is measured to be
\begin{displaymath}
{\cal B}_{\rm MI}(z>0.2) = (0.877 \pm 0.087_{\rm stat} \pm 0.017_{\rm syst}) \times 10^{-6}.
\end{displaymath}
The measurements performed separately for $K^+$ and $K^-$ decays are consistent:
\begin{displaymath}
{\cal B}^+_{\rm MI}(z>0.2) = (0.881 \pm 0.107_{\rm stat}) \times 10^{-6},~~~
{\cal B}^-_{\rm MI}(z>0.2) = (0.868 \pm 0.147_{\rm stat}) \times 10^{-6}.
\end{displaymath}
This is the first published measurement of the $K^-_{\pi\gamma\gamma}$ decay rate.

The observed decay spectrum agrees with the ChPT expectations. The values of the $\hat{c}$ parameter in the framework of the ChPT ${\cal O}(p^4)$ and ${\cal O}(p^6)$ parameterizations~\cite{da96} have been obtained from log-likelihood fits
to the data $z$ spectrum:
\begin{eqnarray}
\hat c_4 &=& 1.37 \pm 0.33_{\rm stat} \pm 0.14_{\rm syst},\nonumber\\
\hat c_6 &=& 1.41 \pm 0.38_{\rm stat} \pm 0.11_{\rm syst}.\nonumber
\end{eqnarray}
Both ${\cal O}(p^4)$ and ${\cal O}(p^6)$ descriptions are equally favoured by the data. These measurements are in agreement with the earlier results reported from $K_{\pi\gamma\gamma}$ decays ($\hat c_4=1.6\pm0.6$, $\hat c_6=1.8\pm0.6$)~\cite{ki97} and $K_{\pi\gamma ee}$ decays ($\hat c_6=0.90\pm0.45$)~\cite{ba08}, and
are obtained at improved precision. The model-dependent branching fraction in the full kinematic range is obtained by integrating the ChPT ${\cal O}(p^6)$ differential decay rate~\cite{da96} for the above value of $\hat c_6$:
\begin{displaymath}
{\cal B}_6(K_{\pi\gamma\gamma}) = (0.910 \pm 0.072_{\rm stat} \pm 0.022_{\rm syst}) \times 10^{-6},
\end{displaymath}
in agreement with an earlier measurement ${\cal B}_6(K_{\pi\gamma\gamma})=(1.1\pm0.3_{\rm stat}\pm0.1_{\rm syst})\times 10^{-6}$~\cite{ki97}. This result also agrees with a prediction for the total decay rate $\Gamma(K_{\pi\gamma\gamma})=76~{\rm s}^{-1}$~\cite{se72} which, considering a mean $K^\pm$ lifetime of $\tau_K=(1.2380\pm0.0021)\times 10^{-8}$~s~\cite{pdg}, translates into ${\cal B}(K_{\pi\gamma\gamma})=\tau_K\Gamma(K_{\pi\gamma\gamma}) = (0.941\pm0.002)\times 10^{-6}$.



\section*{Acknowledgements}

We express our gratitude to the staff of the CERN laboratory and the technical staff of the participating universities and laboratories for their efforts in the operation of the experiment and data processing. We thank Giancarlo d'Ambrosio and Jorge Portol\'es for numerous discussions of the model-dependent analysis.




\begin{thebibliography}{99}
%
\bibitem{se72}
L.M. Sehgal, Phys. Rev. {\bf D6} (1972) 367.
%
\bibitem{ec88}
G. Ecker, A. Pich and E. de Rafael, Nucl. Phys. {\bf B303} (1988) 665.
%
\bibitem{da96}
G. D'Ambrosio and J. Portol\'es, Phys. Lett. {\bf B386} (1996) 403.
%
\bibitem{ge05}
J.-M. G\'erard, C. Smith and S. Trine, Nucl. Phys. {\bf B730} (2005) 1.
%
\bibitem{ki97}
P. Kitching {\it et al.}, Phys. Rev. Lett. {\bf 79} (1997) 4079.
%
\bibitem{ba08}
J.R. Batley {\it et al.}, Phys. Lett. {\bf B659} (2008) 493.
%
\bibitem{fa07}
V. Fanti {\it et al.}, Nucl. Instrum. Methods {\bf A574} (2007) 433.
%
\bibitem{ba07}
J.R. Batley {\it et al.}, Eur. Phys. J. {\bf C52} (2007) 875.
%
\bibitem{geant}
GEANT detector description and simulation tool,\\CERN program library
long writeup W5013 (1994).
%
\bibitem{pdg}
J. Beringer {\it et al.} (Particle Data Group), Phys. Rev. {\bf D86} (2012) 010001.
%
\bibitem{ga06}
C. Gatti, Eur. Phys. J. {\bf C45} (2006) 417.
%
\bibitem{ch67}
N. Christ, Phys. Rev. {\bf 159} (1967) 1292.
%
\bibitem{dafne}
G. D'Ambrosio, M. Miragliuolo and P. Santorelli, DA$\Phi$NE physics handbook, LNF-92/066.
%
\bibitem{ba10}
J.R. Batley {\it et al.}, Eur. Phys. J. {\bf C68} (2010) 75.
%
\bibitem{ga99}
F. Gabbiani, Phys. Rev. {\bf D59} (1999) 094022.
%
\bibitem{ci12}
V. Cirigliano {\it et al.}, Rev. Mod. Phys. {\bf 84} (2012) 399.
%
\bibitem{bi03}
J. Bijnens, P. Dhonte and F. Persson, Nucl. Phys. {\bf B648} (2003) 317.
%
\bibitem{james}
F. James, Statistical Methods in Experimental Physics, 2nd edition, World Scientific, 2006.
%
%
%
\end{thebibliography}
\end{document}